\documentclass[twocolumn]{aastex631}
\usepackage{amsmath}

\shorttitle{Spheric Harmonic Analysis Of Solar Magnetic Power Spectrum}
\shortauthors{Yukun Luo et al.}
\graphicspath{{./}{figures/}}
\usepackage{threeparttable}
\usepackage{booktabs}

\begin{document}

\title{The Sun's Magnetic Power Spectra Over Two Solar Cycles. \uppercase\expandafter{\romannumeral1}. Calibration Between SDO/HMI And SOHO/MDI Magnetograms}

\author{Yukun Luo}
\affiliation{School of Space and Environment, Beihang University, Beijing, People’s Republic of China}
\affiliation{Key Laboratory of Space Environment Monitoring and Information Processing of MIIT, Beijing, People’s Republic of China}

\author{Jie Jiang}
\affiliation{School of Space and Environment, Beihang University, Beijing, People’s Republic of China}
\affiliation{Key Laboratory of Space Environment Monitoring and Information Processing of MIIT, Beijing, People’s Republic of China}

\author{Ruihui Wang}
\affiliation{School of Space and Environment, Beihang University, Beijing, People’s Republic of China}
\affiliation{Key Laboratory of Space Environment Monitoring and Information Processing of MIIT, Beijing, People’s Republic of China}



\begin{abstract}

The Sun's magnetic field is strongly structured over a broad range of scales. The magnetic spatial power spectral analysis provides a powerful tool to understand the various scales of magnetic fields and their interaction with plasma motion. We aim to investigate the power spectra using spherical harmonic decomposition of high-resolution SOHO/MDI and SDO/HMI synoptic magnetograms covering three consecutive solar cycle minima in a series of papers. As the first of the series, we calibrate and analyze the power spectra based on co-temporal SDO/HMI and SOHO/MDI data in this paper. For the first time, we find that the calibration factor $r$ between SOHO/MDI and SDO/HMI varies with the spatial scale $l$ of the magnetic field, where $l$ is the degree of a spherical harmonics. The calibration factor satisfies $r(l)=\sqrt{-0.021 l^{0.64}+2} \quad(5<\mathrm{l}\leq539)$. With the calibration function, most contemporaneous SOHO/MDI and SDO/HMI magnetograms show consistent power spectra from about 8 Mm to the global scales over about 3 orders of magnitudes. Moreover, magnetic power spectra from SOHO/MDI and SDO/HMI  maps show peaks/knees at $l\approx120$ corresponding to the typical supergranular scale (about 35 Mm) constrained from direct velocimetric measurements. This study paves the way for investigating the solar-cycle dependence of supergranulation and magnetic power spectra in subsequent studies.

\end{abstract}



\section{Introduction} \label{sec:intro}

The solar magnetic field ubiquitously distributed on the solar surface is highly structured and multiscale \citep{Solanki2006}. Magnetic features have spatial scales from the global axial dipole field, to active regions (ARs), to the network and internetwork magnetic fields, and further to scales smaller than can currently be resolved \citep{deWijn2009}. The well-organized magnetic features result from the direct and continuous interaction of the magnetic fields with turbulent convective flows. For instance, the supergranulation velocity field is regarded to be responsible for the web-like structure of the photospheric magnetic fields, i.e., network magnetic fields \citep{Leighton1962, Simon1964, Rincon2018}. 

The kinetic and magnetic spatial power spectra over a wide range of scales provide an effective way to explore the physical processes at the origin of different magnetic features and their interaction with plasma flows \citep{Nakagawa1973,Nakagawa1974,Petrovay2001}. Many efforts have been taken to analyze the velocity maps for the kinetic power spectra. Using spherical harmonic decomposition, \cite{Hathaway2000,Hathaway2015} obtain photospheric kinetic energy spectra covering the degree of spherical harmonics, $l$, from $l$=1 to $l\sim$3000 based on the high-resolution full-disk Dopplergrams of MDI onboard the SOHO and HMI onboard the SDO, respectively. Both spectra show distinct peaks representing granulation of about 1 Mm and supergranulation of about 35 Mm. Since supergranulation is strongly linked to the structure of the magnetic field, \cite{Williams2011} compare supergranular characteristics between the minima of cycles 22/23 and 23/24 using the same data and method as \cite{Hathaway2000}. Using the same method, \cite{Williams2014} further analyze Doppler data from contemporaneous HMI and MDI observations. They show that supergranulation appears to be smaller within HMI data than is measured with MDI.

Being different from studies on kinetic power spectra covering a wide range of scales, most previous attempts on magnetic power spectra cover narrower spatial scales, especially one of the two ends of the resolved scales. The constantly improving spatial resolution of magnetic field measurements leads to magnetic spectra studies in the upper end of the spectra covering granular and subgranular scales. Using Fourier decompositions of local magnetograms of the quiet Sun, \cite{Abramenko2001, Goode2010, Stenflo2012, Danilovic2016, Abramenko2020} studied magnetic power spectra aiming to understand properties of internetwork magnetic field and its possible origin, especially its connection with the small-scale dynamo \citep{Vogler2007,Marschalk2013}. The lower end of the magnetic power spectrum corresponds to the global-scale field. Its properties provide insights into the understanding of the solar cycle and the behavior of the Sun as a star \citep{Vidotto2016, Vidotto2018}. \cite{DeRosa2012, Pipin2018} decomposed full-disk synoptic magnetograms from MDI using spherical harmonic decomposition. Their spectra span from $l$=1 to $l\sim$100 to understand the solar cycle and large-scale dynamo \citep{Karak2014, Charbonneau2020}. Most of the rest of studies of the magnetic power spectra concentrate on the analysis of local magnetograms of ARs for the relationship between flare productivity and the magnetic power index \citep{Abramenko2002,Abramenko2010,Mandage2016} or for the variation of magnetic power spectra during ARs flux emergence \citep{Kutsenko2019}.

The aforementioned efforts on magnetic power spectra indicate that there are very few studies of magnetic power spectra emphasizing supergranulation so far. One of the relevant studies is made by \cite{Katsukawa2012}, whose magnetic power spectra show weak peaks at the supergranular scales based on Hinode SOT magnetograms between 2006 November and 2007 December. Given the association between supergranulation and the magnetic network, it is natural to wonder about the following questions. What is the reliable magnetic power spectrum in the broad spatial range covering the supergranular scale? Do the spectra show peaks or knees at the supergranular scale? How do the power spectrum and the scale corresponding to supergranulation vary within a solar cycle and over multiple solar cycles? To our knowledge, all of the questions have not been directly investigated as of now. Many studies on the cycle-dependence of the supergranular size have been carried out by other data sets, e.g., Ca II K images \citep{McIntosh2011} and Doppler velocity data \citep{DeRosa2004}, which give quite controversial results. 

Thanks to successive high-resolution full-disk magnetic field measurements from MDI and HMI since 1996 onwards, we have an opportunity to investigate the magnetic power spectra over a broad range of scales from the global scale to supergranular scale and to a few Mm over cycles 23 and 24. This period covers three cycle minima including the notable deep and extended cycle 23/24 minimum \citep{Jiang2013}. Moreover, cycle 24 is the weakest cycle over the past 100 yr. These provide a rare opportunity to deeply investigate the cycle dependence and activity dependence of the magnetic power spectrum and its supergranulation scale. However, MDI and HMI adopt different spectral lines and have different spatial resolutions and data qualities \citep{Liu2012}. So to investigate the property of the magnetic power spectrum over multiple cycles using measurements from the two instruments, the calibration between the co-temporal HMI and MDI magnetograms is the prerequisite.

Two major methods are widely used to compare and calibrate line-of-sight magnetograms and synoptic maps from different instruments. They are the pixel-by-pixel comparison \citep[e.g.,][]{Liu2012,Riley2014,Pietarila2013,Tran2005} and histogram methods \citep[e.g.,][]{Riley2014, Jones1992, Jones2000, Wenzler2006}. The calibration factor varies somewhat with center-to-limb distance and field strength. But neither method can answer whether the calibration factor varies with the spatial scales of the magnetic field. \cite{Virtanen2017} introduce a new method for scaling the photospheric magnetic field in terms of the harmonic expansion and investigate the scaling of the harmonic coefficients between some pairs of data sets (without the pair of MDI and HMI). The method can give an independent scaling factor for the different harmonic terms, corresponding to different spatial scales. The spatial-scale dependence of the calibration factor affects the configuration of the magnetic power spectrum. This inspires us to explore the calibration between MDI and HMI data in terms of the spherical harmonic decomposition as the first step to investigating the properties of the magnetic power spectrum, in this study. 

In summary, this paper commences a series of investigations into the Sun's magnetic power spectrum aiming to quantify the solar cycle dependence of the properties of the magnetic power spectrum over the broadest range of scales so far. The objective of the first paper is twofold: spatial-scale dependent calibration between MDI and HMI synoptic magnetograms and identification of the supergranular scale based on calibrated MDI and HMI co-temporal magnetic power spectra. The second paper will study the variation of the magnetic power spectra after the calibration during 1996-2020 covering cycles 23 and 24 with three consecutive cycle minima. We will put emphasis on the temporal variation of the supergranular size and the power indices for the inertial range between AR scales and supergranular scales in the course of the solar cycle and over 2 cycles. In the third paper, we will focus on the temporal variation of the spectra at the range that is larger than AR scales.

This paper is organized as follows. We describe MDI and HMI synoptic magnetograms and the method of obtaining the magnetic power spectra by spherical harmonic decomposition in Section \ref{sec:datas}. In Section \ref{sec:results}, we present our calibration of co-temporal MDI and HMI magnetograms. The identification and characterization of the supergranulation pattern based on the calibrated magnetic power spectra are also presented in this section. We summarize and discuss the above results in Section \ref{sec:conclusion}.

\section{data and analysis}\label{sec:datas}

\subsection{Data} \label{subsec:data}

In our series of studies, we use two radial synoptic magnetogram data sets covering the 23rd and 24th solar cycles. They are observed by Michelson Doppler Imager (MDI) on board the Solar and Heliospheric Observatory (SOHO) \citep{Scherrer1995} and the Helioseismic and Magnetic Imager (HMI) on board the Solar Dynamics Observatory (SDO) \citep{Scherrer2012,Scherrer2012b}, respectively. The two instruments utilize different spectral lines with various resolutions: the Ni \uppercase\expandafter{\romannumeral1} 6768 \AA \ (Fe \uppercase\expandafter{\romannumeral1} 6173 \AA) \citep{Norton2006} is adopted as the spectral line for the MDI (HMI), with coverage of $3600\times1080 \ (3600\times1440)$ pixels. The data sets begin in Carrington Rotation (CR) 1911 (July 1996) and end in CR 2225 (December 2019, data missing for CRs 1938-1942 and CR 1945). In this paper, we focus on analyzing the synoptic maps for the overlap period, i.e. from CR 2097 to CR 2104.

The initial synoptic maps lack data on the polar magnetic fields, which are crucial for the decomposition of the spherical harmonic function. Therefore, we use polar-corrected data \citep{Sun2011,Sun2018}. Specifically, they interpolate and smooth the north (south) polar magnetic fields by using a multi-year series of well-observed polar fields from each September (March). This process enables synoptic maps with polar field correction to be obtained. It is worth noting that the polar correction approach of HMI differs slightly from that of MDI, and the implications of these differences are analyzed in Section \ref{subsec:polar}.

\subsection{Spherical harmonic decomposition and power spectra}
\label{subsec:spherical}

For the magnetic field $B(\theta,\phi)$, we can express it by the following Eq. (\ref{eq1})
\begin{equation}\label{eq1}
B(\theta ,\varphi )=\sum_{l=0}^{\infty }\sum_{m=-l}^{l}  B_{lm} Y_{lm}(\theta ,\varphi ) ,
\end{equation}
where $\theta$ is the colatitude, $\phi$ is the longitude, $Y_{l,m}(\theta ,\varphi )$ is the spherical harmonic function with degree $l$ and azimuthal order $m$, and $B_{l,m}$ is the corresponding spherical harmonic coefficient.

\begin{figure}
\plotone{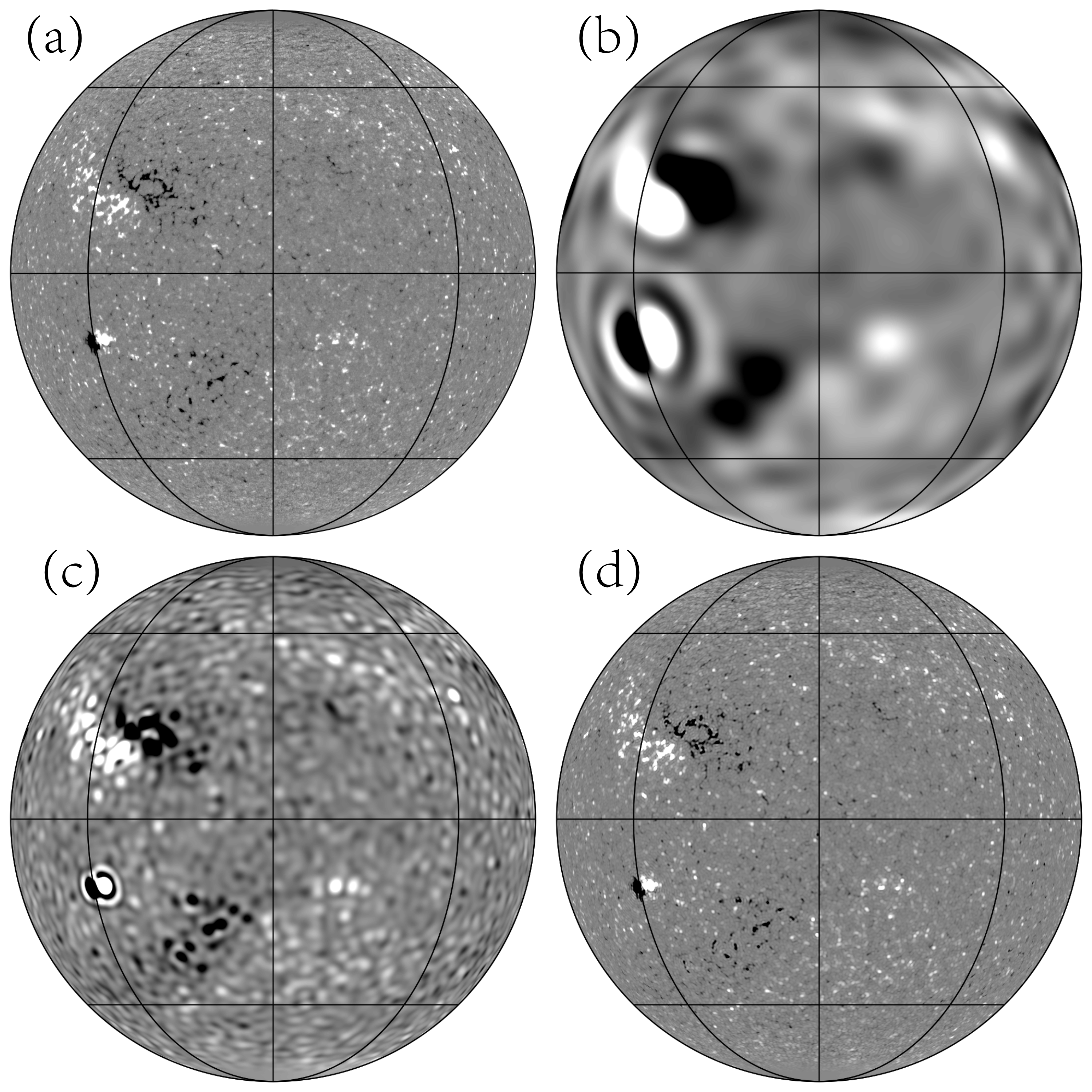}
\caption{CR 2097 synoptic map as an example of the spherical harmonic expansion. An orthographic projection of the map centered at the equator and the longitude of 0$^{\circ}$ is adopted. (a) the original synoptic map; (b)-(d) spherical harmonics reconstruction up to the maximum degree $l_{max} =30, 120$, and 539, respectively. The magnetic fields in Panels (a)-(d) are saturated at 60, 8, 25, and 60 G, respectively.}
\label{fig:spheric}
\end{figure}

For each synoptic map, we use the ``pyshtools" package in Python to perform a spherical harmonic decomposition. The algorithm is proposed by \cite{Driscoll1994}, who employ a grid suitable for data with different sampling densities at different latitudes and introduce weight coefficients. For example, the longitudinal distances of the pixel points near the equator are larger in synoptic maps than those near the polar region. Therefore, the data need to be weighted when performing the spherical harmonic decomposition. The discretization formula for obtaining the spherical harmonic coefficient for a grid of size $2n\times2n$, introducing the weighting factor $a_j$, is given by:
\begin{equation}\label{eq4}
    B_{lm}=\frac{\sqrt{2\pi } }{2n}  \sum_{j=0}^{2n-1}  \sum_{k=0}^{2n-1}a_jB(\theta_j,\phi_k )Y_{lm}(\theta_j,\phi_k ), 
 \end{equation}

where $\theta_j = \pi j/2n$, $\phi_k = \pi k/n$, corresponding to spherical coordinates for any point on the sphere, and the number of grid points in both latitude and longitude is $2n$. The weight coefficient $a_j$ is given by the following equation
\begin{equation}\label{eq5}                                     
\begin{aligned}
    a_{j}=\frac{2 \sqrt{2}}{n} \sin \left(\frac{\pi j}{\mathrm{n}}\right) \sum_{l=0}^{\frac{n}{2}-1} \frac{1}{2 l+1} \sin \left([2 l+1] \frac{\pi j}{n}\right),\\
    \mathrm{j}=0, ……, \mathrm{n}-1.
\end{aligned}
\end{equation}

This algorithm requires the grid to be $n\times2n$ or $n\times n$. Therefore, as the first step in data processing, we transform the data resolution to $2160\times1080$ for MDI and $2880\times1440$ for HMI by a simple linear two-point interpolation algorithm. Afterward, we also convert data from an equal sine-latitude distribution to an equal latitude distribution. At this step, we first calculate the sine of latitude at equal intervals. Then the same two-point interpolation algorithm is used to transform the data based on sine-latitudes. After these processes, the data can be used for ``pyshtools".

Due to the instrumental resolution limitations and sampling theorem, we decompose the data up to $l=539$ and $l=719$ for MDI and HMI synoptic maps, respectively. In the subsequent analysis, we consider only spherical harmonic decomposition results of $l<540$ for both data sets. As an example, Figure \ref{fig:spheric} presents the magnetic field components at different spatial scales after decomposing the CR 2097 synoptic map, which is shown in Figure \ref{fig:spectra} (a). Figure \ref{fig:spheric} (a) is the spherical projection of the original synoptic map. Panels (b)-(d) show maps reconstructed by spherical harmonics up to different maximum degrees. Large-scale structures become progressively more refined from Panels (b) to (d) with $l_{max}$ increasing from 30 to 539. These expected results illustrate the algorithm's validity for the spherical harmonic decomposition of the magnetograms. We can therefore use the results to obtain the power spectra.

Using the orthogonality between the spherical harmonic functions, Parseval's theorem in the Cartesian coordinate system can be extended to the spherical geometry. That is,
\begin{equation}\label{eq6}
    \int_{\Omega} B^{2}(\theta, \phi) d \Omega=\sum_{l=0}^{\infty} P(l),
\end{equation}

where $P$ is the power spectrum and it is related to the spherical harmonic coefficients by
\begin{equation}\label{eq7}
    P(l) =\sum_{m=-l}^{l}B_{lm} ^2.
\end{equation}

It is clear from Eq. (\ref{eq7}) that in the power spectrum, information about the scale of azimuthal order $m$ is combined. In our work, the spectral power represents the magnetic energy at the spatial scale corresponding to the spherical harmonic degree $l$. To convert the degree $l$ to the actual spatial scale $\lambda$, we can use the following equation:
\begin{equation}\label{eq8}
    \lambda=\frac{2 \pi R_{\odot}}{\sqrt{l(l+1)}},
\end{equation}
where the parameter $R_{\odot}$ is the solar radius.

\section{results} \label{sec:results}

\subsection{Scale dependence of MDI \& HMI data}\label{subsec:scale}

\begin{figure*}[ht]
\plotone{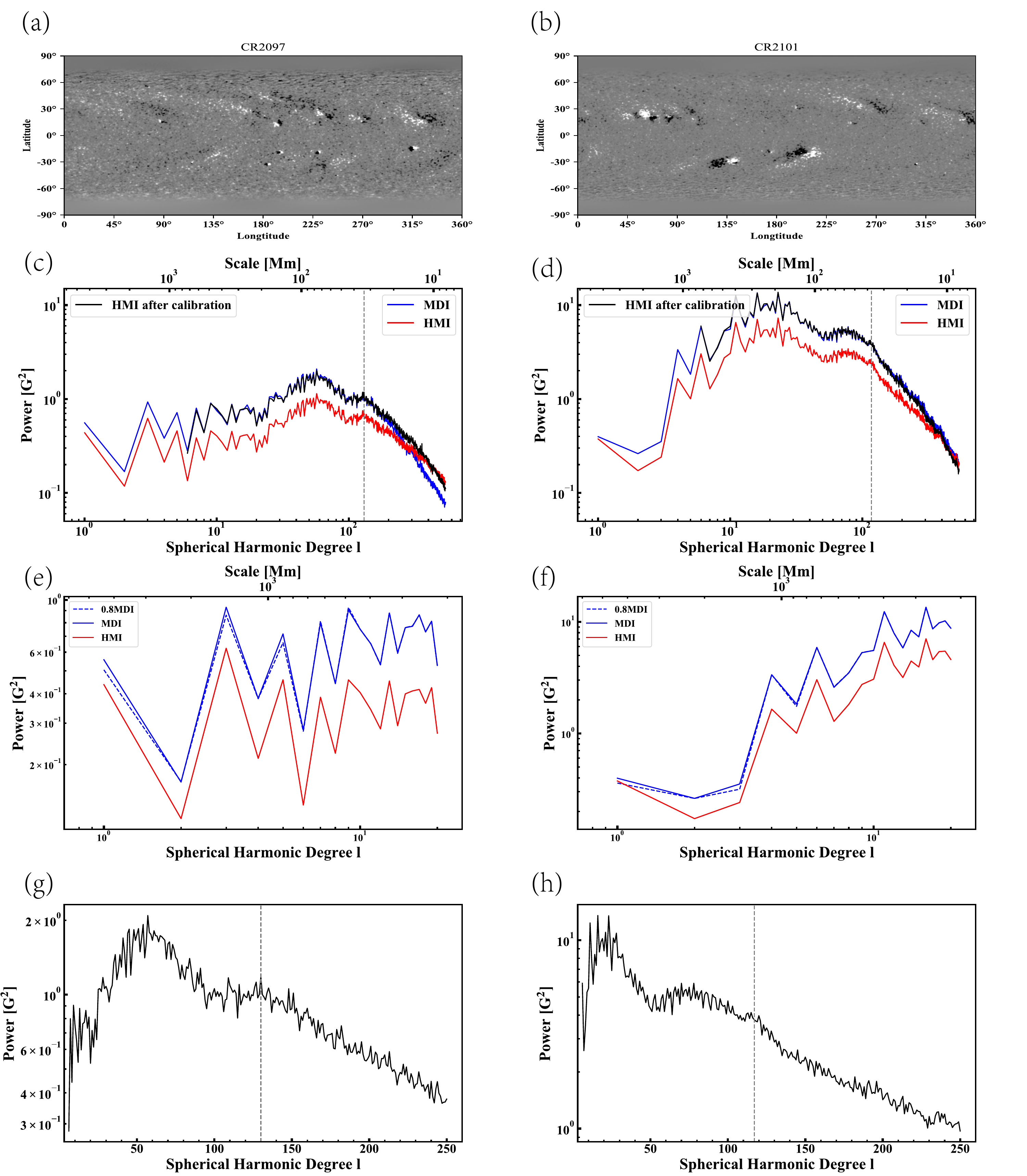}
\caption{Magnetic power spectra obtained by MDI and HMI synoptic magnetograms for CR 2097 (left) and CR 2101 (right). (a)-(b) Synoptic maps, magnetic fields are saturated at 53 G and 100 G, respectively. (c)-(d) Power spectra of the HMI map (red), MDI map (blue), and calibrated HMI map (black). (e)-(f) Power spectra for $l_{max}=20$. The blue dotted lines are the power spectra of MDI maps multiplied by a factor of 0.8 on polar fields. (g)-(h) Results of supergranule identification based on the spectrum from $l$=6 to $l$=250. The location of the peak or knee is marked with vertical gray dotted lines.}
\label{fig:spectra}
\end{figure*}

After the above data processing steps, we obtain the magnetic power spectra of eight CRs during the overlapping period. As examples, Figures \ref{fig:spectra} (a) and (b) show the synoptic magnetograms for CR 2097 and CR 2101, respectively, while the corresponding HMI and MDI magnetic power spectra are shown in Figures \ref{fig:spectra} (c) and (d). The same figures but for the remaining six CRs are shown in the appendix. 

Figures \ref{fig:spectra} (c) and (d) show that both magnetic field strength and spatial scales influence the difference between HMI and MDI power spectra, which is consistent with previous studies \citep{Liu2012, Riley2014, Pietarila2013, Virtanen2017}. We next investigate whether the impact of these two factors is minimal enough to use a fixed calibration factor or whether it is necessary to calibrate based on spatial scales. 

\begin{figure}
\plotone{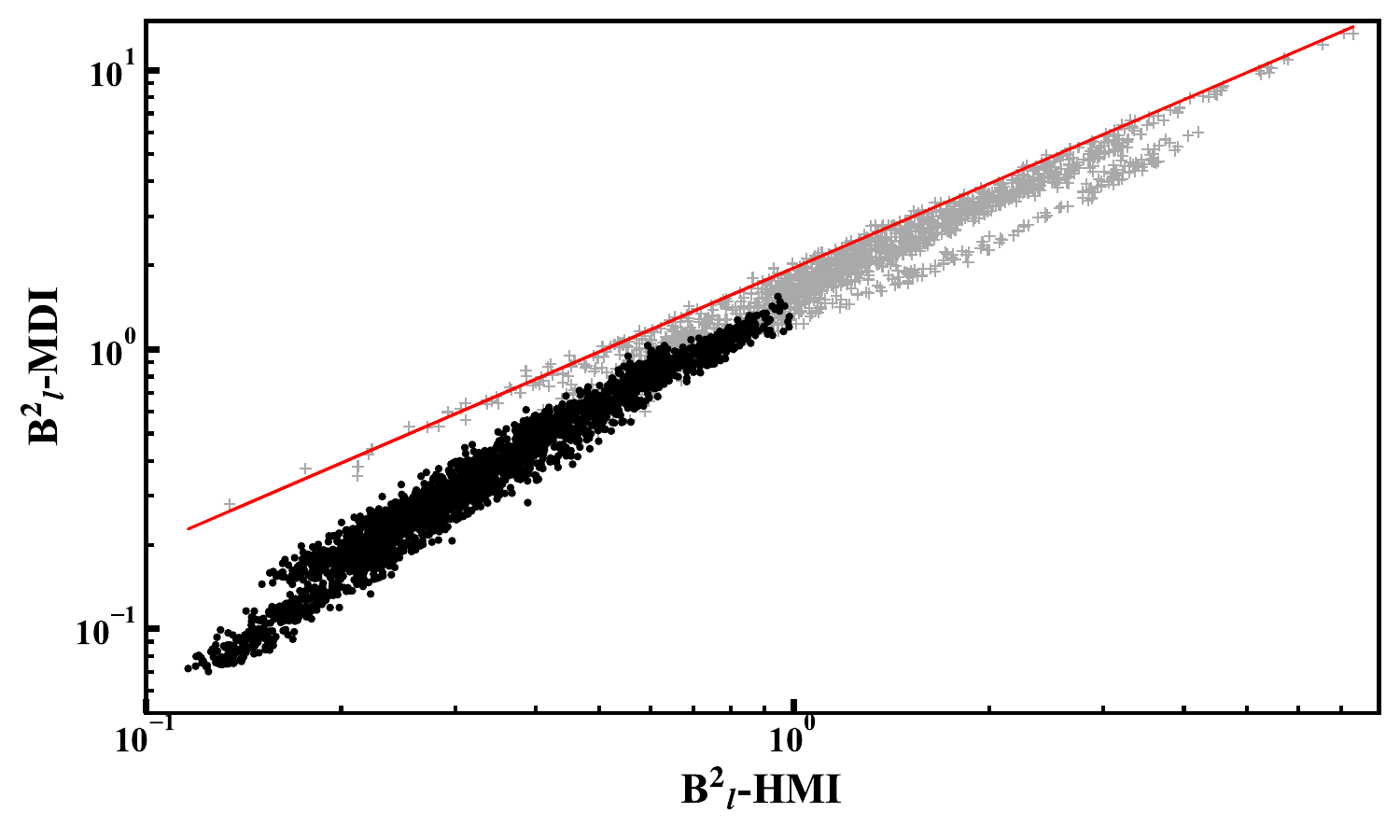}
\caption{Comparison of HMI and MDI spectral power. The horizontal and vertical axes are the spectral power of HMI and MDI, respectively. The gray and black dots correspond to spherical harmonical degrees $l<200$ and $l>200$ respectively. The red line represents the fixed calibration factor 1.4, i.e. $B^2_{l-MDI}=1.96B^2_{l-HMI}$ (1.4 for $B_l$).}
\label{fig:1.4}
\end{figure}

We present a comparison of the MDI and HMI spectral power for data during the overlapping period in Figure \ref{fig:1.4}. The red line represents the calibration results obtained by \cite{Liu2012}. The calibration factor, $1.4$, corresponds to the vertical intercept in the log-log plot. The plot shows that the distribution of points is non-linear, with a significant deviation from the red line. That is, the calibration factor varies enough to make a fixed calibration factor inapplicable. Thus we need to develop a new calibration method applicable to power spectra analysis.

Notably, black dots in Figure \ref{fig:1.4} (small-scale part of magnetic fields) generally have smaller calibration factors and lower spectral power than grey dots (large-scale part). This may suggest that the calibration factors are smaller for weaker magnetic fields. However, this seems to contradict the findings of \cite{Liu2012}, which indicate that the calibration factors are smaller for stronger magnetic fields. This discrepancy may arise from the stronger influence of spatial scale on the calibration factor than that of magnetic strength, with smaller spatial scales corresponding to smaller factors. These results indicate scale dependency in the difference between MDI and HMI magnetograms. Therefore, it is essential to consider the calibration at various scales in our work.

\subsection{Influence of polar magnetic field on the power spectra}\label{subsec:polar}

In the theory of spherical harmonic decomposition, the term $l=0$ represents the average result of the global scale magnetic field. For the magnetic field, the term should be zero theoretically. However, this coefficient is not zero based on observed magnetograms, which may be introduced by noise or during the construction of the synoptic magnetograms \citep{Bertello2014}. We will not consider this term in the analysis that follows. The power spectra in this paper all start from $l=1$.

In Section \ref{subsec:data}, we mention that \cite{Sun2011, Sun2018} use similar approaches to correct the polar magnetic field of the HMI and MDI synoptic magnetograms. However, the approaches require the use of the previous year's data for interpolation. But HMI synoptic magnetograms for the overlap period do not have the previous year's data. So for this part, they use the data of MDI instead. In addition, the polar fields are smoothed and weighted with the non-polar data in the correction. These correction methods increase uncertainties of the polar magnetic field and further influence the magnetic power spectra. Therefore, we need to determine the calibration range according to the range of the power spectra affected by the polar field.

We multiply the data above 75\textdegree \ north and south of MDI magnetograms by a factor of 0.8 to test the effect of the polar field. We then obtain their power spectra, which are shown in Figures \ref{fig:spectra} (e) and (f). It can be seen that the scaled MDI spectral power corresponding to the dash line has a certain decrease at $l=1,3,5$ and are close to HMI. This just supports the idea that the polar magnetic field has an impact on the power spectra of the large scale field ($l\leq5$).
The results are similar to \cite{Virtanen2016}: The correction of the polar field affects axial dipole ($l=1$) and quadrupole ($l=2$) but has no effect on the results of a smaller scale. Thus, the effect of the polar field uncertainty for $l>5$ can be ignored.
Taking these results together, we will calibrate the power spectra for $l>5$ and the calibration results will be applied in the non-overlap period.

\subsection{Scale dependent calibration function for MDI and HMI magnetograms}\label{subsec:calibration}

\begin{figure*}[ht]
\plotone{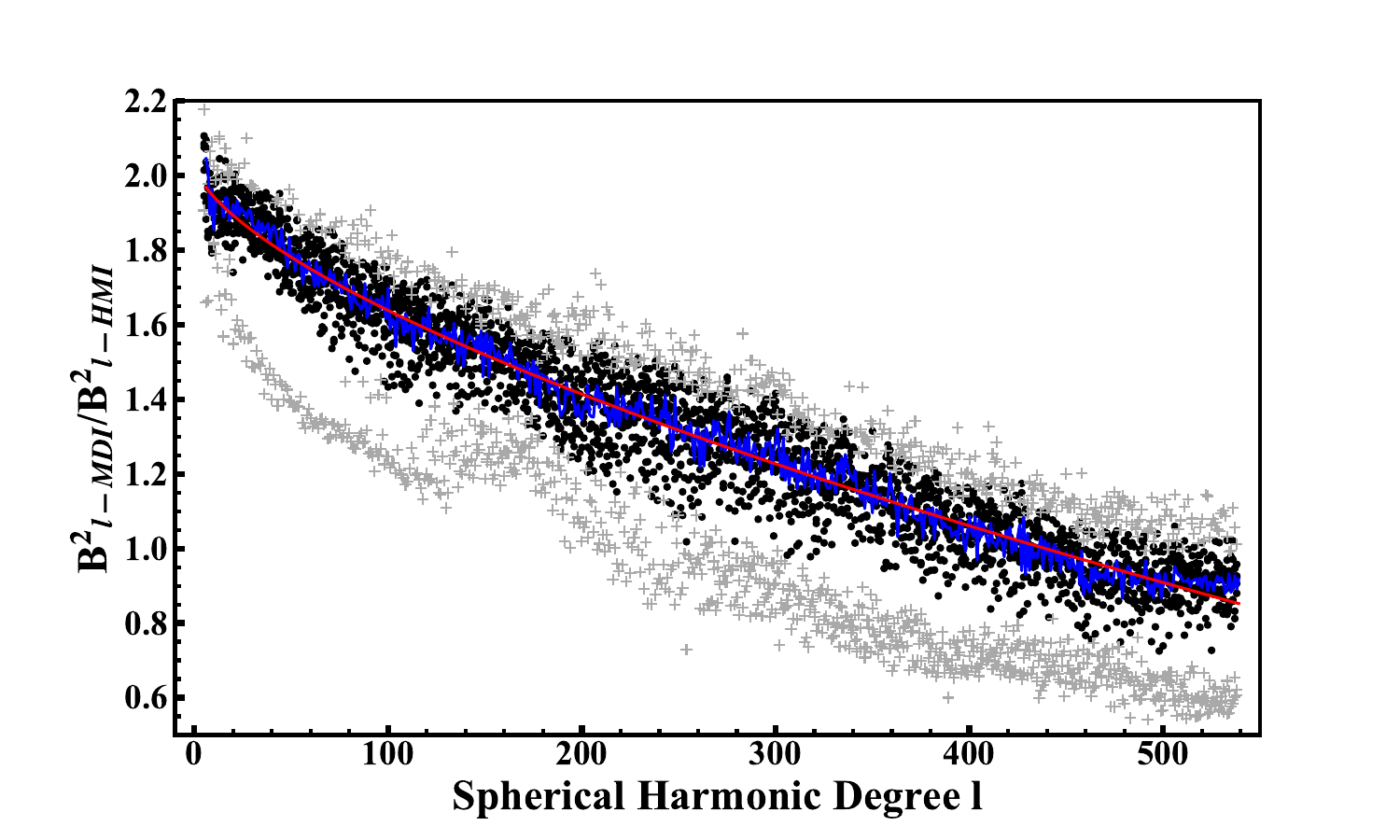}
\caption{Spatial scale dependency of the calibration factor between MDI and HMI spectral power. Black and gray dots are the observation data within and outside $\sigma$, the variance of the calibration factors for a given $l$, respectively. The blue line is the mean of the black dots and is fitted to give the scale-dependent calibration function $r(l)$ (red line).}
\label{fig:function}
\end{figure*}

\begin{table*}[ht]
\caption{Quantitative analysis and comparison of calibration results}
\label{tablec}
\centering
\begin{threeparttable}
\begin{tabular}{cccccc}
\hline \hline
CR   & Consistent interval of $l$\tnote{a} & $RSS(r=1)$\tnote{b} & $RSS(r=1.4)$\tnote{b} & $RSS(r(l))$\tnote{b} & $\Delta B_{r(l)}$/$\Delta B_{r=1.4}$ \\ \hline
2097 & 6-251               & 42.23    & 32.07     & 2.26     & 0.806     \\
2098 & 6-178               & 24.73   & 48.67     & 5.55    & 0.810     \\
2099 & 6-539               & 374.25    & 87.21     & 2.84    & 0.822     \\
2100 & 6-539               & 163.74   & 29.12     & 4.19    & 0.812     \\
2101 & 6-539               & 893.03   & 121.02     & 4.32    & 0.785     \\
2102 & 6-539               & 318.19   & 91.14     & 3.68    & 0.800     \\
2103 & 6-539               & 169.69   & 75.51     & 1.63    & 0.819      \\
2104 & 162-539             & 153.16   & 436.56     & 145.14    & 0.807     \\ \hline
\end{tabular}
\begin{tablenotes}
\footnotesize
\item[a] Consistent interval is the range for $l$ where MDI and calibrated HMI power spectra are consistent. 
\item[b] Residual sum of squares between the MDI and HMI power spectrum calibrated with the calibration factor $r$=1 (uncalibrated one), $r$=1.4, and calibration function Eq.(\ref{eq9}), respectively.
\end{tablenotes}
\end{threeparttable}
\end{table*}

In the first two subsections, we have shown that the calibration factors are scale-dependent in the range $l>5$. To investigate this further, we combine data for the eight CRs with $l>5$ and explore the pattern of variation of the spectral power ratio $B^2_{l-MDI}/B^2_{l-HMI}$ with the spherical harmonic degree $l$. The result is shown in Figure \ref{fig:function}. The plot clearly shows a decreasing trend in the ratio as the degree $l$ increases and this trend is consistent across different CRs. This further suggests that the calibration factor is a function of $l$.

Due to the limited number of CRs with the same $l$, the impact of noise or measurement error is relatively large. To mitigate the impact, we exclude special points by discarding data larger than the variance $\sigma$ for a given $l$. The remaining data points are then averaged to obtain the solid blue line in Figure \ref{fig:function}. We use a nonlinear function to fit the averaged data and obtain the following calibration function $r(l)$: 
\begin{equation}\label{eq9}
    r(l)=\frac{B_{l-MDI}}{B_{l-HMI}}=\sqrt{-0.021 \ell^{0.64}+2.02}.
\end{equation}
The calibration factor $r$ for $l=5\ (\approx800$Mm) is 1.4, while for $l=30\ (\approx140$Mm) and $l=140\ (\approx30$Mm), the calibration factors are 1.35 and 1.23, respectively. The factors are close to the calibration factor of 1.4 obtained in \cite{Liu2012}. These demonstrate that the calibration factors obtained through traditional methods for analyzing full-disk magnetograms are mostly relevant for large-scale structures, such as active regions. Our results suggest that the classical methods may have limited applicability to smaller-scale features. In contrast, \cite{Riley2014} found a calibration factor $r\approx1.05$ for the CR 2097 synoptic magnetogram. This value is smaller than our calibration factor for the large-scale section but larger than that for the small-scale section. As shown in Figure \ref{fig:spectra} (c), small-scale features contribute significantly to the total power, which suggests that the overall single calibration factor obtained by averaging our calibration factors may be similar to the value of 1.05. 

In addition, \cite{Virtanen2017} concluded that the calibration factors for low spherical degrees are smaller than those obtained through pixel-by-pixel comparison or histogram techniques when they performed spherical harmonic decomposition for synoptic maps. However, it should be noted that their work did not involve the calibration of HMI and MDI data sets, thus the applicability of their conclusion to the calibration of these two data sets is uncertain. Hence, we should be careful to directly compare our findings with theirs.

\subsection{Evaluation of the new calibration method}\label{subsec:analysis}

After applying the calibration function Eq. (\ref{eq9}) to the power spectra, we first evaluate the similarity between the calibrated HMI and MDI power spectra. Specifically, we consider the calibrated HMI and MDI power spectra to be consistent if the difference between them is less than $20\%$ of the MDI spectral power. We list the consistent interval of $l$ based on this criterion in the second column of Table \ref{tablec}. Our calibration method leads to throughout consistent HMI and MDI power spectra for CRs 2099-2103, based on this criterion. Moreover, the power spectra of CR 2097 and CR 2098 exhibit consistency only on the large-scale part, from $l=6$ to around $l\approx200$. And the consistent interval for CR 2104 is mainly in the small-scale part.

The exception of CR 2104 is also apparent in the corresponding images. In Figure \ref{fig:1.4}, below the red line, there is a cluster of gray dots, which results from CR 2104. In Figure \ref{fig:function} gray points below the black dots show a significant deviation. This part of the data is also from CR 2104. These results suggest that there might be some issues with the CR 2104 HMI or MDI magnetograms, which could be attributed to instrumental measurements.

To illustrate in more detail the applicability of our calibration method for the power spectra analysis, we additionally use two quantitative comparisons. The residual sum of squares ($RSS$) measures the difference between the MDI power spectra and the uncalibrated ($r=1$) or calibrated HMI power spectra. The formula is as follows.
\begin{equation}\label{eq10}
    RSS=\sum_{l=6}^{539}\left(B^{2}_{l-HMIc}-{B}^{2}_{l-MDI}\right)^{2}.
\end{equation}
The results are shown in columns 3 to 5 of Table \ref{tablec}, where $r=1.4$ represents the fixed calibration factor 1.4 and $r(l)$ represents the calibration function Eq. (\ref{eq9}). 
The smaller the RSS value is, the smaller difference between the two power spectra is. While using a fixed calibration factor, the RSS is reduced to only one-fifth to one-half of the uncalibrated case. For CR 2098 and CR 2104, the RSS even shows an increase. However, when using the calibration function $r(l)$, all the RSS show a decrease, with the vast majority reduced to a tenth or even a hundredth of the uncalibrated case. This shows that our calibration method is more suitable for power spectra analysis.

In addition to using the RSS to evaluate the effect of our calibration on the power spectra, we also quantify the effect on the magnetograms. To accomplish this, we perform an inverse spherical harmonic decomposition of the power spectra to reproduce the magnetograms. Using the MDI magnetograms as the reference, we separately obtain the HMI magnetogram using the two calibration methods (fixed calibration factor 1.4 and calibration function $r(l)$). The ratio of the difference between the HMI and MDI magnetograms at a given pixel is obtained using the following formula:
\begin{equation}\label{eq11}
    \frac{\Delta B_{r(l)}}{\Delta B_{r=1.4}}=\frac{B_{MDI}-B_{HMI-r(l)}}{B_{MDI}-B_{HMI-r=1.4}}.
\end{equation}
For each CR map, we then conduct a linear regression of the results for all pixels. The results are summarized in the last column of Table \ref{tablec}. Notably, the ratios of magnetogram differences for all CRs are within a narrow range of $0.81\pm0.01$. These quantitative comparisons indicate the efficacy of our calibration approaches in achieving not only a better analysis of power spectra but also a more consistent between HMI and MDI synoptic maps.

\subsection{Supergranule identification}\label{subsec:supergranulation}

\begin{table*}[ht]
\centering
\caption{Location of peak or knee identified on power spectra of MDI and calibrated HMI synoptic magnetograms for CRs 2097-2104.}
\label{tables}
\begin{threeparttable}
\begin{tabular}{ccccccc}
\hline \hline
CR   & $l_{sgr1}(HMI)$ & $l_{sgr1}(MDI)$ & $l_{sgr2}(HMI)$ & $l_{sgr2}(MDI)$ & $l_{sgr3}(HMI)$ & $l_{sgr3}(MDI)$ \\ \hline
2097 & 130        & 130        & 131±5      & 132±5      & 129      & 130      \\
2098 &            &            &            &            & 138      & 133        \\
2099 & 140        & 140        & 156±5      & 156±5      & 143      & 139      \\
2100 & 123        & 123        & 128±5      & 128±5      & 127      & 127      \\
2101 & 123        & 120        & 119±5      & 119±5      & 117      & 117      \\
2102 &            &            &            &            & 133        & 132        \\
2103 &            &            &            &            &            &            \\
2104 &            &            &            &            & 125        & 125        \\ \hline
\end{tabular}
\begin{tablenotes}
\footnotesize
\item \textbf{Notes.} The symbols `$l_{sgr1}$', `$l_{sgr2}$', and `$l_{sgr3}$' are degrees of spherical harmonics of the identified peak or knee corresponding to supergranulation using the three methods presented in Section \ref{subsec:supergranulation}.
\end{tablenotes}
\end{threeparttable}
\end{table*}
The analysis of Dopplergrams using spherical harmonic functions, as discussed in Introduction, shows the peak around $l=120$ on the kinetic power spectra corresponding to supergranulation. In Section \ref{subsec:calibration} we have mentioned that MDI and calibrated HMI power spectra are consistent around this degree. So it is feasible to search for supergranule features in the magnetic power spectra.

Based on Figure \ref{fig:spectra} and the power spectra images in the appendix, it is apparent that a majority of the magnetic power spectra exhibit peaks or knees around $l=120$. Here we use three algorithms to verify the existence of these structures and the corresponding specific spatial scales. We start with a morphological approach based on the code developed by \cite{Duarte2021} to find the peaks in the spectra. We search and identify peaks by the trend of the curve around the local maximum, the threshold, and the width. By adjusting these parameters, the identification results can be optimised. The corresponding results are in columns 2 and 3 ($l_{sgr1}$) of Table \ref{tables}. The second method is similar to wavelet analysis, where a wavelet is fitted to the peak and the presence of a peak is discerned based on the result of fit \citep{Du2006}. This method is achieved through the `scipy.signal.find\_peaks\_cwt' within Python. Since the method does not give the peak, but gives the center of the wavelet, we list the range of the peak rather than the exact value, with the results in columns 4 and 5 ($l_{sgr2}$) of Table \ref{tables}. The third method we use is the algorithm for finding peaks and knees used in the field of neural power spectra: FOOOF \citep{Donoghue2020}. This method gives information on the peaks of the four CRs around the supergranule scale and also gives the knee of the three CRs. The results are in columns 6 and 7 ($l_{sgr3}$) of Table \ref{tables}.

The supergranulation identification results are between $l=116 \ (36$ Mm) and $l=140 \ (30$ Mm), which are consistent with the results in the kinetic power spectra. Notably, for the same CR, we observe slight differences in the results obtained from different methods. For instance, in the case of CR 2099, the first two methods yield a difference of $\Delta l=16$. The third method, meanwhile, can assist in determining which result is more reliable. Therefore, by combining the three methods, we can identify the supergranule scale correlated features of the power spectra of most CRs. However, we are unable to identify significant features in the power spectra of CR 2103, possibly due to the portion of power spectra belonging to active regions masking supergranulation.

Furthermore, in the work of \cite{Williams2014}, they observe different sizes of supergranulation in the kinetic power spectra for HMI and MDI, with $l=132$ for HMI and $l=119$ for MDI. They attribute the difference to the different spatial resolutions. They show that by smoothing the resolution of HMI Dopplergrams, a power spectrum similar to MDI data can be obtained. However, our identifications do not indicate any significant differences between the HMI and MDI results, which remain generally consistent.

Regarding the absence of differences between the supergranular scales in our results mentioned above, we first exclude the effect of the calibration. We note that the difference of calibration factors ($\Delta r=0.02$) around the supergranule scale cannot significantly alter the scale of the feature. Furthermore, the identified supergranule sizes in HMI and MDI maps of CR 2104 remain consistent, even though their power spectra do not coincide in $l<162$. We suggest that the difference in the identification results is due to the influence of small-scale structures. \cite{Hathaway2002,Rincon2017} mention that the kinetic power spectra near the size of supergranulation can be affected by granulation, which are not completely removed. The higher the resolution, the more features of granulation are included. Therefore, smoothing maps with high resolution is essentially a way to obtain similar power spectra by reducing the influence of small-scale structures. Magnetic power spectra in this paper are obtained from synoptic magnetograms, which have a relatively low resolution. Multiple full-disk magnetograms are averaged in the construction of a synoptic map. Strong small-scale magnetic structures and noise are removed. Therefore, we can obtain ideal results in the magnetic power spectra. On the other hand, in addition to small-scale structures, large-scale structures such as active regions may also cause an unphysical shift of the position of supergranulation in the power spectra. We need to be careful about this potential effect in future work.

\section{conclusion and discussion} \label{sec:conclusion}

In this work, we conduct spherical harmonic decomposition of HMI and MDI synoptic magnetograms for their overlapping period to obtain magnetic power spectra. A scale-dependent calibration function between MDI and HMI spectral power $r(l)=\sqrt{-0.021 \ell^{0.64}+2.02}$ for magnetic power spectra analysis is derived. In some magnetic power spectra, the peak or knee around $l=120$ associated with supergranulation is clearly presented. This study lays the foundation for further research on the cycle dependence of supergranulation. 

The scale-dependent calibration factors might be mainly attributed to the different effects of magnetic filling factors on the measurement of magnetic fields with different scales. The magnetic filling factor is set to the constant value of 1 when both the MDI and HMI magnetograms were derived from polarimetric measurements \citep{Berger2003, Liu2012, Centeno2014, Grinon-Marin2021}. As shown in Figure \ref{fig:spheric}, the larger-scale magnetic field is more uniformly distributed. For the smaller-scale field having distinct magnetic structures, lower-resolution data tend to give lower magnetic field strength since only the mean flux density contributes to the Stokes profiles. Hence we see the scaling factor $r(l)$ decrease with $l$ when we scale higher-resolution HMI data to lower-resolution MDI data. In addition, the different modulation transfer functions for the two instruments could be the other ingredient contributing to the scale-dependent calibration factors \citep{Abramenko2001, Stenflo2012}, especially near the resolution limit of the MDI maps, i.e., around the largest $l$. The scale-dependent calibration factors have been illustrated by \cite{Virtanen2017}, who investigated different pairs of datasets. But note that they scaled lower-resolution data to higher-resolution one, and the scaling factors typically increase with $l$. There is no particular reason for us to choose the lower-resolution data, MDI synoptic maps, as the reference in this paper.

In contrast to a constant calibration factor, our proposed calibration function exhibits superior suitability for power spectra analysis while simultaneously reducing the discrepancies between magnetograms after calibration. Furthermore, as highlighted in \cite{Virtanen2017}, the calibration approach is independent of the resolution of magnetograms, thereby reducing the potential impact of resolution on the calibration results. Despite the limited number of synoptic magnetograms available during the overlap period of MDI and HMI, our analysis of the power spectra of these magnetic maps reveals a consistent and scale-dependent variation of the calibration factors, with no apparent signs of temporal variation. We limit our scale-dependent calibration between HMI and MDI synoptic magnetograms to the range $l>5$ because of the unreliable measurement of the polar field. The calibration function could be extended to the range $l\leq5$, which gives $r$ about 1.4. The value is essential for the understanding of the polar field distribution and long-term evolution of the solar open flux \citep{Jiang2013,linker2017}.

So far there are very few studies of supergranulation-scale magnetic fields with power spectra of MDI or HMI synoptic maps. For the first time, to the best of our knowledge, we derive the magnetic power spectrum over the broad spatial scale from about 20 Mm to the global scale and identify supergranulation based on the magnetic spectra. Although what we detected is actually the magnetic network, rather than the supergranulation flow, the magnetic network has a direct association with supergranulation \citep{Rincon2018}. 

Most magnetic power spectra we investigated show a distinct peak or knee at the supergranular scale of about 35 Mm ($l\approx120$), which is consistent with the typical size measured from the Doppler velocity observations. The consistency is expected given the association between supergranulation and the magnetic network and the consistency also is a piece of evidence to support the association. In contrast, some previous investigations on the scale of supergranulation using different methods show divergent results. For example, applying local correlation tracking techniques to Doppler velocity data of MDI, \cite{DeRosa2004} derived the average supergranular cell diameter lies in the 12–20 Mm range. \cite{Meunier2008} show the average size is in the range of 16-17 Mm. However, the supergranulation scale is not necessarily a distinct feature on all magnetic power spectra as presented in Table 2 since active regions tend to disorganize and wash away supergranulation \citep{Hindman2009}. Due to the effects of active regions and other sources of noise, the identification of the peak or knee of the power spectra is slightly method-dependent. The slightly different locations of the peak or knee correspond to slightly different sizes of supergranulation. Besides this random source for the variation of supergranular sizes, there may also exit an inherent variation of supergranular sizes with solar activity \citep{Meunier2007}.

Knowledge of the variation of the supergranular size presented by the magnetic power spectrum is crucial for a better understanding of the interaction between the magnetic field and convection at different scales and of the solar total and spectral irradiance variations \citep{Yeo2014}. In our subsequent work, we will refine our identification methods of supergranulation from magnetic power spectra and extend our analysis to the entire solar cycles 23 and 24. We will investigate how the size of supergranulation varies within a solar cycle and over multiple cycles.\\ \hspace*{\fill} \\

We thank the referee for the valuable comments. The research is supported by the National Natural Science Foundation of China No. 12173005 and National Key R\&D Program of China No. 2022YFF0503800. J.J. acknowledges the International Space Science Institute Teams 474. We would like to express our gratitude to the teams responsible for the development of Python toolkits such as `pyshtools' and `scipy'. The SDO/HMI data are courtesy of NASA and the SDO/HMI team. SOHO is a project of international cooperation between ESA and NASA.  

\clearpage
\appendix
\begin{figure*}[h]
\plotone{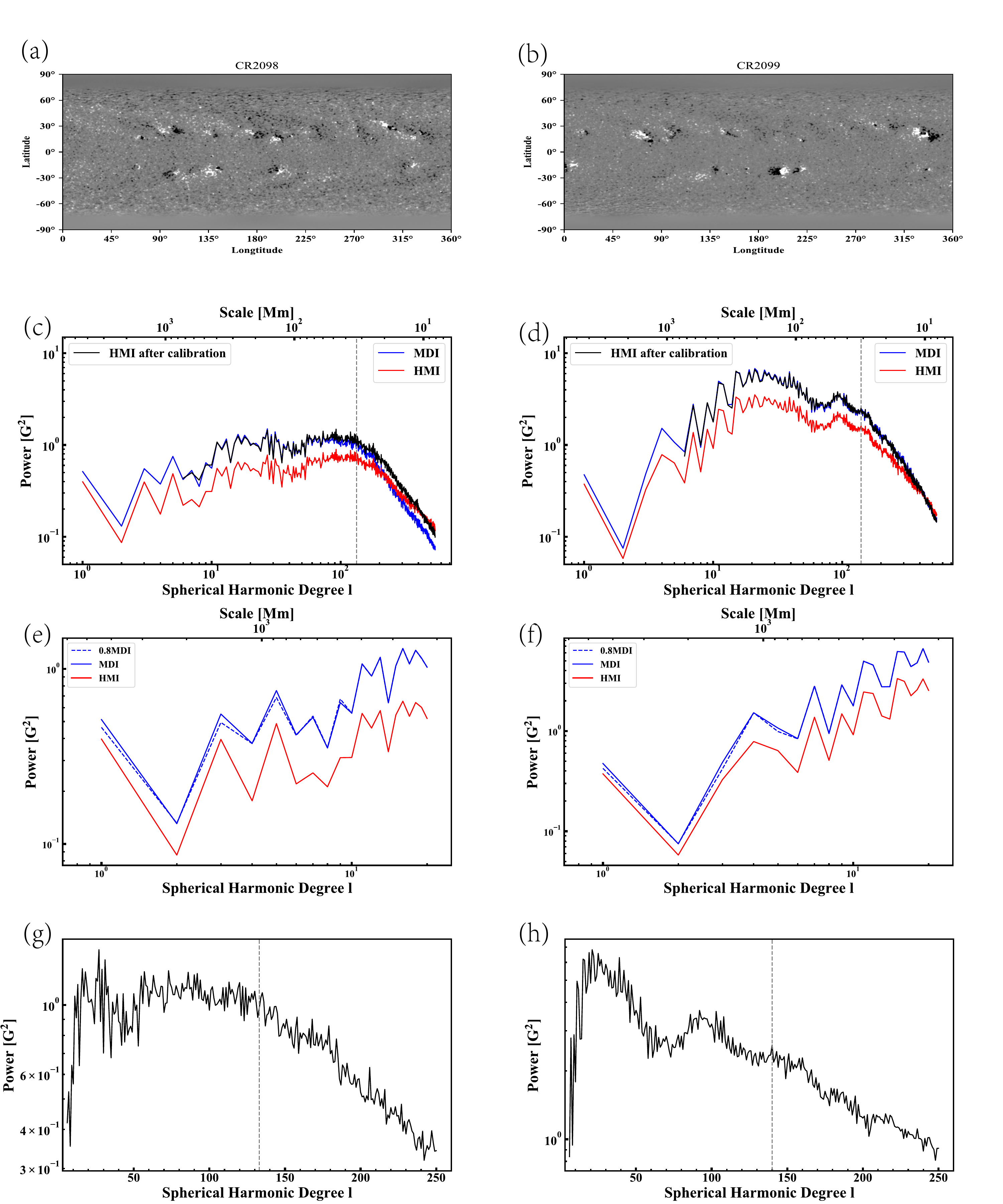}
\caption{Same as Figure \ref{fig:spectra}, but for CR 2098 (left) and CR 2099 (right).}
\label{fig:spectra1}
\end{figure*}

\begin{figure*}[h]
\plotone{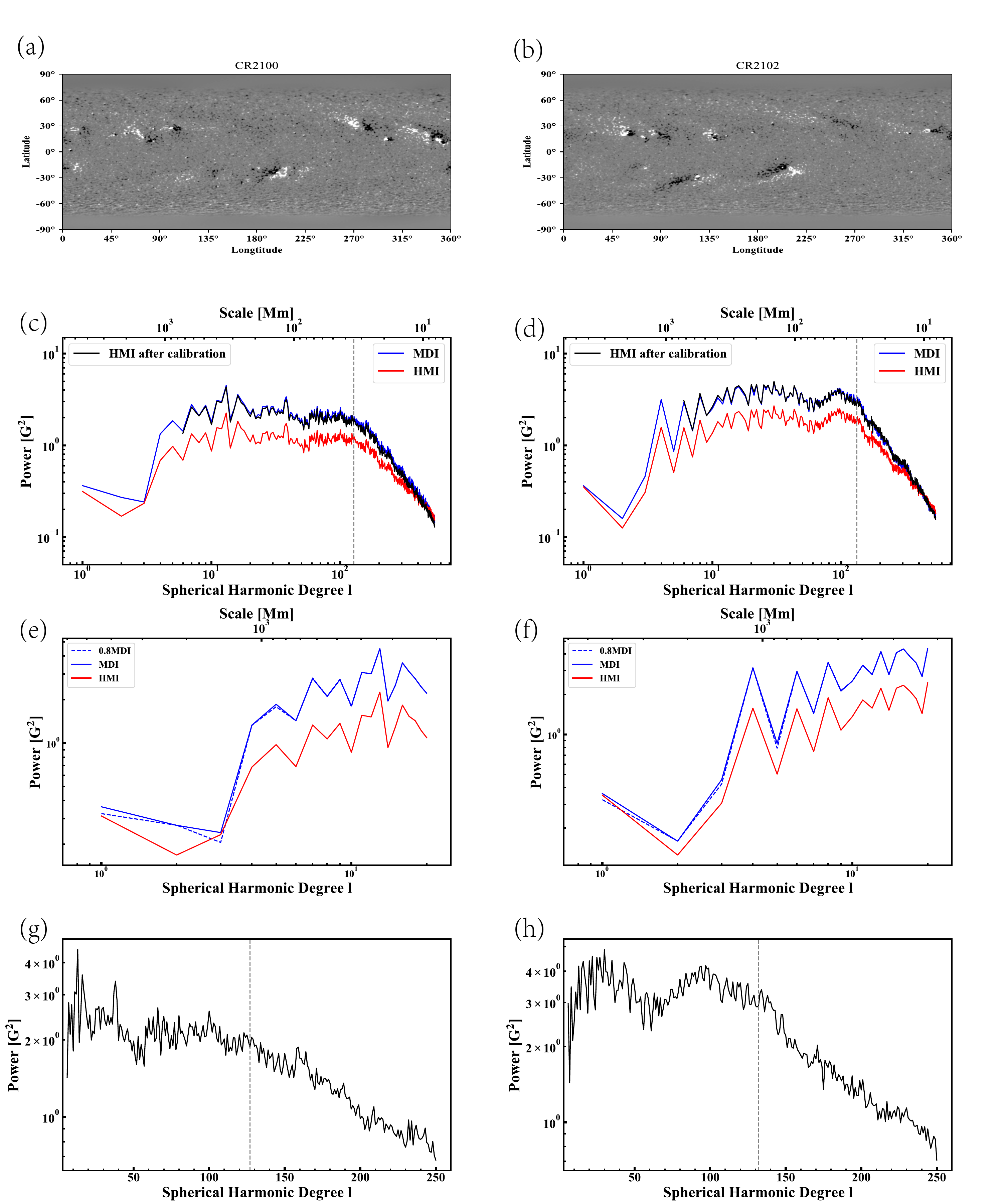}
\caption{Same as Figure \ref{fig:spectra}, but for CR 2100 (left) and CR 2102 (right).}
\label{fig:spectra2}
\end{figure*}

\begin{figure*}[h]
\plotone{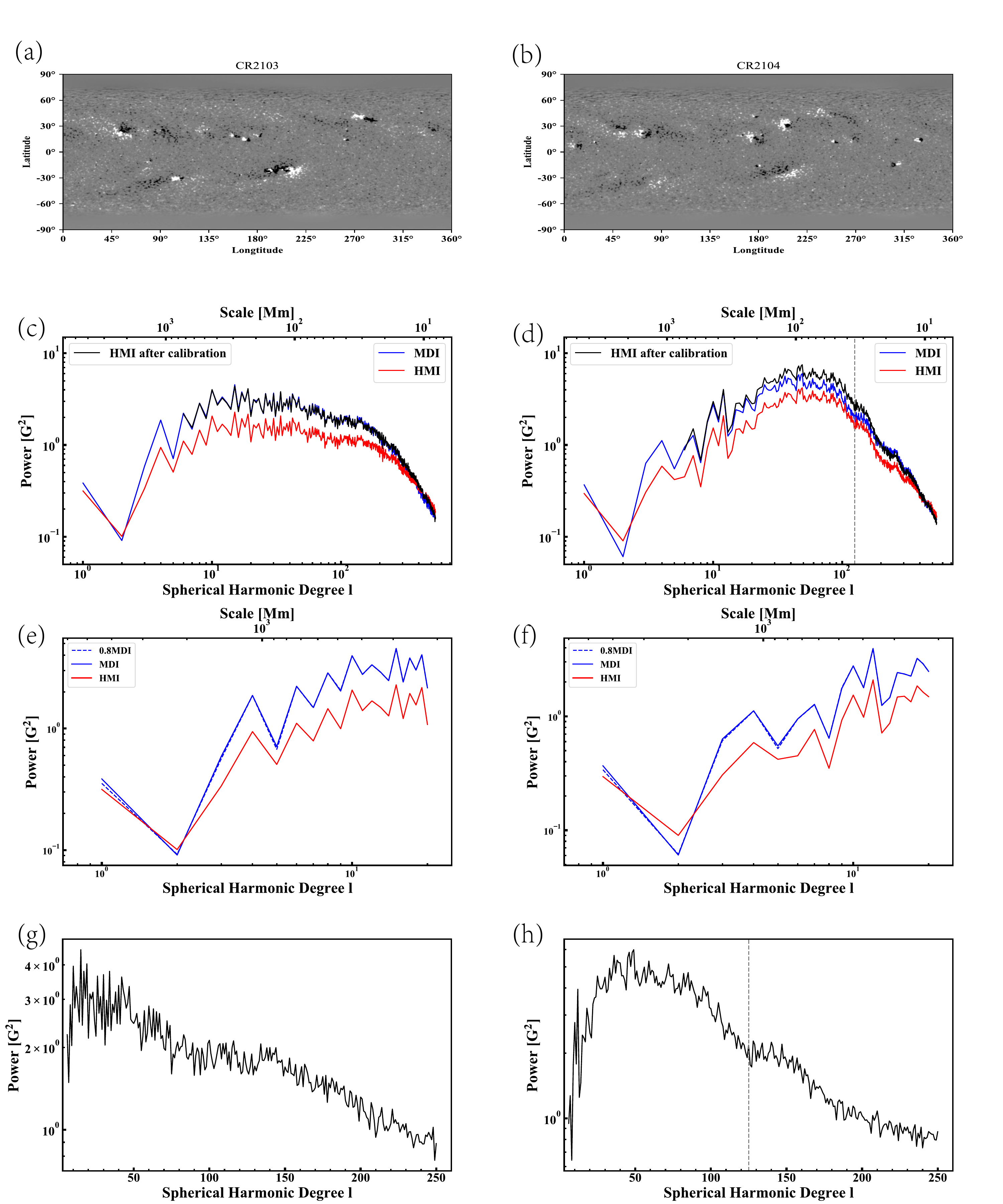}
\caption{Same as Figure \ref{fig:spectra}, but for CR 2103 (left) and CR 2104 (right).}
\label{fig:spectra3}
\end{figure*}
\clearpage
\bibliography{power_spectra}{}
\bibliographystyle{aasjournal}



\end{document}